\documentstyle[epsf]{aipproc}

\renewcommand{\theequation}{\arabic{equation}}
\newcommand{\be}{\begin{equation}}
\newcommand{\ee}{\end{equation}}
\newcommand{\bea}{\begin{eqnarray}}
\newcommand{\eea}{\end{eqnarray}}

\renewcommand{\theequation}{\arabic{equation}}

\begin{document}

\title{\hfill $\mbox{\small{
$\stackrel{\rm\textstyle SU-4240-681\quad}
{\rm \textstyle \quad}
$}}$ \\[1truecm]
The Quantum Hall effect, Skyrmions and Anomalies}
\author{A. Travesset}
\date{}
\maketitle
\thispagestyle{empty}
\begin{center}
{\it Department of Physics, Syracuse University,\\ 
Syracuse, New York 13244-1130,  USA}
\end{center}

\begin{abstract}

We discuss the properties of Skyrmions in the Fractional Quantum  
Hall effect (FQHE). We begin with a brief description of
the Chern-Simons-Landau-Ginzburg description of the FQHE, which provides
the framework in which to understand a new derivation of the properties 
of FQHE Skyrmions 
(S. Baez, A.P. Balachandran, A. Stern and A. Travesset {\em cond-mat 9712151})
from anomaly and edge considerations.

\end{abstract}

\bigskip
\bigskip

\section{The Quantum Hall effect}

\begin{figure}[htb]
\epsfxsize=3in \centerline{\epsfbox{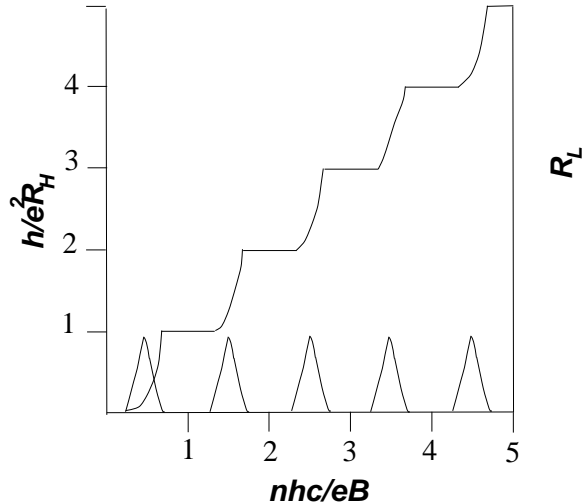}}
\caption{The plateaus in the integer Hall effect resistivity plotted versus
the filling fraction. The left vertical axis is the inverse transverse
resistivity in dimensionless units, and the right one the longitudinal
in arbitrary units. With higher resolution settings smaller plateaus at 
fractional filling fractions may be observed as well. }
\label{fig__experim}
\end{figure}

Experiments carried out starting at the end of the 70's 
(see \cite{Review}) revealed that in some samples, the Hall effect 
presents fascinating and completely unexpected properties, the most 
outstanding being the broad plateaus for the transverse conductivity 
$\sigma_H$ (or for the transverse resistivity $R_H=1/\sigma_H$)
and the vanishing of the longitudinal resistivity at 
this plateaus, see Fig.~\ref{fig__experim}. There are other important 
properties, such as the incompressibility of the Hall liquid and the 
presence of quantum mechanically induced currents if the sample has edges. 
It is fair to say that, in spite that there are still open questions, 
theorist have succeeded in giving successful approaches to the problem.
We give an introduction to the most common field theory formulation 
\cite{zhkl}, but we will necessarily be very descriptive. We refer,
for example, to \cite{SHOU} for a pedagogical review and \cite{NORW} 
for a recent 
and more detailed analysis. The second part of the talk is devoted to the
properties of FQHE Skyrmions, a problem
which has already been addressed in this MRST meeting \cite{soto}.

\subsection{The CSLG Field theory of the FQHE}

A very intuitive picture to understand the FQHE is to consider the
total flux in the sample $\Phi=BA$,
where $B$ is the external magnetic field and $A$ the total area,
as being carried by $N_f$ particles, the fluxons, each one 
carrying an elementary flux unit $\Phi_0=\frac{2\pi\hbar c}{e}$
\footnote{unless explicitly stated, $\hbar=c=1$ units are used.}.
The physical electrons are represented as composite objects, a bare 
bosonic particle with an odd number of fluxons attached to it
restoring its fermionic nature.
The filling fraction is defined as
\be\label{def_fill}
\nu=\frac{N_{el}}{N_f} \ ,
\ee
where $N_{el}$ is the number of physical electrons in the sample.
If $\nu=\nu_k\equiv \frac{1}{2k+1}$, $k \in {\bf N}$, 
there are exactly $2k+1$ fluxons per electron. It is, therefore, an
``optimal'' situation. In any other filling fraction 
\footnote{We do not consider generalized filling fractions.} there 
will be an unbalanced number of fluxons with respect to electrons.
How the system responds to this situation is the key for understanding
the properties of the QH liquid.

The previous picture may be described in a Landau-Ginzburg formalism, the
Chern-Simons-Landau-Ginzburg (CSLG) Lagrangian density being \cite{zhkl}
\bea\label{lag_CSLG}
{\cal L}&=&\phi^{\dagger}({\bf r})[i\partial_t-ea_0]\phi({\bf r})-
\frac{1}{2m^*}\left|\phi^{\dagger}({\bf r}) \left[-i\nabla-e({\bf A}({\bf r})+
{\bf a}({\bf r})) \right]\phi({\bf r})\right|^2-
\nonumber\\\nonumber
&&-\frac{1}{2}\int d {\bf r^{\prime}} V({\bf r -r^{\prime}})
[|\phi({\bf r})|^2-\rho_0][|\phi({\bf r^{\prime}})|^2-\rho_0]
-\frac{1}{2}g \mu_B B \phi^{\dagger}({\bf r}) \sigma_z
\phi({\bf r})- 
\nonumber\\
&&-\frac{e^2}{4\vartheta}\epsilon^{\mu \nu \sigma} 
a_{\mu}({\bf r})\partial_{\nu} a_{\sigma}({\bf r})\ .
\eea
First of all $\phi=(\phi_1,\phi_2)$ is a two-component complex bosonic
field (the ``bare electron''). The field ${\bf a}({\bf r})$ is a 
gauge field that attaches $\frac{\vartheta}{\pi}=(2k+1)$ fluxons
to the boson, which makes the physical electron. $\rho_0$ is the 
actual density of electrons (which obviously coincides with the density
of bosons).  The external magnetic field is described 
by ${\bf A}$ ($\nabla \times {\bf A}= -Be_{z}$). There are two extra terms,
$V({\bf r})=\frac{e^2}{\epsilon |{\bf r}|}$ accounts for the Coulomb repulsion 
between electrons, and the coupling
between the spin and the magnetic field, which involves $g^s$, the
gyromagnetic ratio, the Bohr magneton $\mu_B$ and the Pauli matrix
$\sigma_z$.

The Landau-Ginzburg ground state is the lowest energy solution of 
Eq.~\ref{lag_CSLG}. Let us try the ansatz $\phi=\sqrt{\rho_0}(1,0)$. 
The equations of motion imply
\be\label{eq_motion}
\frac{N_{el}}{A}=\rho_0=\frac{B}{\Phi_0}\frac{\pi}{\vartheta}=\frac{1}{2k+1}
\frac{N_{f}}{A} \ .
\ee
So, the optimal situation described above, in which each electron captures
an odd number fluxons ($2k+1$ of them) is the ground state. The filling
fraction is then $\nu=\nu_k=\frac{1}{2k+1}$,  the longitudinal resistivity
is zero and the transverse conductivity $\sigma_H=\frac{\nu_k e^2}{2 \pi}$.
The case $g^{S}\rightarrow 0$ is special, since the solution is not unique 
as $\phi=\sqrt{\rho_0}(1,0)U$, $U \in SU(2)$ is a solution of the 
equations of motion. However, the true ground state
is ferromagnetic, i.\ e.\ corresponds to a unique choice of $U$.
The $SU(2)$-symmetry is then spontaneously broken to $U(1)$.
the ground state remains spin polarized at $g^S=0$ as well 
(although spins do not necessarily point to the same direction of the 
magnetic field). As a consequence, there are
Goldstone modes in the system manifesting in the form of
neutral disipationless spin waves(magnons). 
But, remarkably, the properties of the ground state are essentially 
independ of $g^{S}$.

To go further, let us temporally simplify Eq.~\ref{lag_CSLG} and neglect the
Coulomb repulsion. Let us add the term
\be\label{vort_core}
-\frac{\lambda}{2}(\phi^{\dagger}({\bf x})\phi({\bf x})-\rho_o)^2 \ .
\ee
If $B$ is the external magnetic field at the ground state, we find solutions 
to the equations of motion in the form 
\bea\label{vort_qhe}
\phi_1(r,\theta)&=&g_V(r)e^{-i\theta}
\nonumber\\
\phi_2(r,\theta)&=&0
\nonumber\\
a_{r}(r,\theta)&=&0
\nonumber\\
a_{\theta}(r,\theta)&=&\frac{eB}{2}r+n_V(r)
\nonumber\\
a_{0}(r,\theta)&=&m_V(r)
\eea
where $lim_{r \rightarrow \infty} g_V(r)=\sqrt{\rho_0}$, 
$lim_{r \rightarrow 0} g_V(r)=0$ and 
$lim_{r \rightarrow \infty} n_V(r)=\frac{N_{vort}}{r}$.
This solution has finite energy and corresponds to a vortex of vorticity 
$N_{vort}$ located at $r=0$ with a finite core $\xi \sim\sqrt{\lambda}$. 
The addition of the term Eq.~\ref{vort_qhe} determines the size of the vortex.
The vortex may be regarded as a particle on its own, the 
pseudo-particle, having electric charge and spin. As an example,
we compute the total electric charge in the ground state of a QH 
liquid in the presence of a vortex. The equations of motions imply
\be\label{motion1}
J^0_{em}=\frac{e}{2\vartheta}\epsilon^{0 \nu \sigma}\partial_{\nu} a_{\sigma}
\ee
So, the total electric charge is
\be\label{total_charge}
Q=\int_{\Sigma} d^2{\bf r} J^{0}_{em}({\bf r})=
\frac{e}{2 \vartheta} \oint_{\partial \Sigma} dl a_l=
-e(N_{el}-\nu_k N_{vort}) \ ,
\ee
A vortex has charge $\frac{e}{2k+1}$. It is then a {\em fractionally} 
charged object. Its spin may be 
also be computed with result $\frac{N_{vort}}{2k+1}$. So, in general,
FQHE vortices are {\em anyons}. 

Now, already at this stage, we can explain some of the phenomenology 
of the FQHE. Let us 
assume that the QH liquid is at its ground state. Each fermion
is a composite boson with an even number of fluxons attached. If the
filling fraction is changed from $\nu_k=\frac{1}{2k+1}$ by 
tuning the external magnetic field for example, the statistical gauge
field ${\bf a}$ screens the excess of magnetic field keeping 
Eq.~\ref{eq_motion} unaltered. That is, the density of electrons $\rho_0$ 
remains constant under small variations of the magnetic field. The 
system is therefore incompressible. Furthermore, as the number of electrons
does not change, the area of the sample $A$ remains constant as well,
a property that is usually referred to as area-preserving.
At some point, however, the magnetic field is enough to create/destroy 
a fluxon. The QH-liquid responds creating a vortex/anti-vortex. 
A vortex(or an antivortex) may 
change dramatically the properties of the liquid. Being a charged 
object, the FQHE properties may be destroyed because of the motion of
those vortices. However, metals with a sufficient amount of disorder
may pin down those vortices, and the properties of the ground state
like the longitudinal and transverse conductivity are not changed. That 
explains the broad plateaus observed.
Eventually, more vortices cannot be accommodated in the 
sample, and the longitudinal conductance grows from its zero
value. This corresponds to a transition to the next 
ground state filling fraction. The CSLG description nicely accounts for
the experimental situation in Fig.~\ref{fig__experim}.

The case $g^{s}\sim 0$ is experimentally relevant in some cases
such as $GaAs$ samples. We may try solutions to the equations of 
motion in the form 
\bea\label{skyrm_qhe}
\phi_1(r,\theta)&=&g_S^{(1)}(r)e^{-i\theta}
\nonumber\\
\phi_2(r,\theta)&=&g_S^{(2)}(r)
\nonumber\\
a_{r}(r,\theta)&=&0
\nonumber\\
a_{\theta}(r,\theta)&=&\frac{eB}{2}r+n_S(r)
\nonumber\\
a_{0}(r,\theta)&=&m_S(r) \ ,
\eea
where the assymptotics are the same as in 
Eq.~\ref{skyrm_qhe}, but in addition, we require 
$lim_{r \rightarrow 0} g_S^{(2)}(r)\ne 0$,
$lim_{r \rightarrow \infty} g_S^{(2)}(r)=0$. This solution minimizes
the Coulomb energy; when the spin up component starts
to decrease the spin down increases and the electron density deviations 
from the ground state value $\rho_0$ are much smaller than for the vortex. 
The spin contribution increases, but it is anyhow negligible
since the assumed small gyromagnetic ratio. It should not
come as a surprise then, that in this regime, fluctuations in spin have
a lower energy than vortices, and become the relevant 
quasi-particles \cite{Sondhi}. 
those solutions are of a topological nature as well, they are Skyrmions,
as we will see.

Although the ground state is essentially independent of $g^{S}$, 
the properties of the pseudo-particles are very different. There is a critical
value $g_c^{S}$ such that for $g^{S}<g_c^{S}$ those are Skyrmions, while for
$g^{S}>g_c^{S}$ they are vortices. 

We have introduced the pseudo-particles and explained the crucial
role they have in explaining the properties of the QH fluid.
From now on, we concentrate on the study of its properties,
and more specifically of FQHE Skyrmions.  
In the case of vortices, this may be accomplished by integrating out
the bosonic field, i.\ e.\ going to the dual picture \cite{SHOU}.
The properties of vortices, such as the statistics or its electric 
charge, are explicit.

In \cite{WE} we have generalized this description to the case of
Skyrmions. As it is well known, the spin of Skyrmions may be determined
from the coefficient of the Hopf term. We include this term and read
the spin from its coefficient.
To determine the couplings of the different terms in the Lagrangian we
consider a sample with edges. In the presence of edges, there are
quantum mechanically induced chiral gapless modes \cite{Fro}.
We shall determine all the properties
of FQHE Skyrmions by imposing this chirality condition, together with
the fact that Skyrmions may be created and destroyed.

\subsection{The properties of FQHE Skyrmions}

Let us write the direction to which the spin field is pointing at in 
terms of a unitary vector $\vec n$. It is possible to everywhere identify an
$SU(2)$ field degree of freedom $g$, associated with spin fluctuations
from
\be\label{spin_fluc}
n_i=\frac{1}{2}{\rm Tr}(\sigma_i g^{\dagger} \sigma_3 g) \ .
\ee
There is an $U(1)$ subgroup, which we arbitrarily take to be 
generated by the third  Pauli matrix $\sigma_3$, which 
is gauged via the coupling to a statistical gauge
field (the same gauge field mentioned above), so that the gauge invariant
observables are defined on $S^2$ (parametrized by $\vec n$).
Energy finiteness generally demands that $g$ goes to the above 
$U(1)$ subgroup, i.e. $g\rightarrow \exp{i\chi\sigma_3}$,
at spatial infinity.
This corresponds to $\phi^{\dagger}\phi$ going to the ground state
value at spatial infinity, and in effect this compactifies
${\bf R}^2$ to $S^2$. This shows the topological nature of the FQHE Skyrmions.
They are naturally associated with $\Pi_2 (S^2)$,
the elements of $\Pi_2(S^2)$ being labeled by the winding number
\be
N_{Sky}=\int_{{\bf R}^2} d^2x \; T^{0}(g)=
\frac{i}{4\pi}\int_{{\bf R}^2} d\; {\rm
Tr}\;\sigma_3
g^\dagger dg \;, \label{}
\ee 
where $T^0(g)$ is the time component of the topological  current,
\be T^{\mu}(g)=
\frac{i}{4\pi}\epsilon^{\mu\nu\lambda} \; {\rm  Tr}\sigma_3\partial_\nu
g^\dagger \partial_\lambda g \;. \label{topcur}
\ee

We first show what are the consequences of not including a Hopf
term. Our starting point is the standard bulk action of the dual
picture \cite{Fro}, including a coupling to the Skyrmion current
\be\label{Hall_LAG}
  {\cal S}_H=\int_{\Sigma\times R^1} d^3x \;\biggl(
\frac{\sigma_H}{2}\epsilon^{\mu \nu \lambda} A_{\mu}
\partial_{\nu}  A_{\lambda}-  eA_{\mu}{\cal
 T}^{\mu} \biggr) \label{qhlag} \;,
\ee
where $\Sigma$ is the  two dimensional spatial domain of the sample,
$R^1$ accounts for time, and $e{\cal T}^{\mu}$ is the Skyrmion current.
Additional terms  may be added \cite{WE}, but we just
consider  the topological sector that determines the charge and the
spin. For us, $A_\mu$ is the external electromagnetic field which is not
a dynamical variable. Its variations therefore just
define the bulk current $J_{em}^\mu$ by
$ J^\mu_{em} = -  \frac{\delta {\cal S}_H}{\delta A_{\mu}}  .$

According to Eq.~\ref{Hall_LAG}, the bulk electromagnetic current 
$J^{\mu}_{em}$ is
\be\label{OUR_MOT1}     J^\mu_{em}= -
\frac{\sigma_H}2 \epsilon^{\mu \nu \lambda} F_{\nu\lambda}+
e {\cal T}^{\mu} \; .\label{Halleq}
\ee
For consistency  the current $J^\mu_{em}$, and consequently
 ${\cal  T}^\mu$, must be conserved.  This is the case for
 ${\cal  T}^\mu$
 proportional to the topological current $T^\mu$:
\be {\cal T}^\mu =\kappa
 T^\mu \;.\ee
Eq.~\ref{Halleq} implies that the electric charge density is
$J^{0}_{em}=-\sigma_H F_{12}+e \kappa T^{0} $. Integrating it over
the whole sample gives the total electric charge 
as $-eN_{el}+e\kappa N_{Sky}$, where $N_{el}$ is the total number
of electrons at the corresponding filling fraction $\nu=\frac{1}{2k+1}$. 
Thus, $\kappa$ times $e$ is the charge of a Skyrmion of unit winding number.

We now examine under what conditions the bulk action $ {\cal S}_H$ is
consistent with the existence of chiral edge currents.
For the case of filling fraction $\nu=1$, there is a
single edge current on the boundary  $\partial \Sigma$
of $\Sigma$, which may be represented by a 2d massless chiral
relativistic Dirac fermion \cite{Fro},
while for fractional values of $\nu$ one gets a Luttinger liquid.
Chirality implies that the electromagnetic current $J_{em}^\mu$ of the
edge fermions  satisfies
$   J_-^{em} =   \frac{1}{\sqrt{2}}(J^0_{em} + J^1_{em})
 =0 $.  In the quantum theory
this is known to lead to an anomaly, i.e. $ \partial_\mu J^\mu_{em} \ne 0$.

It is convenient to bosonize the edge theory\cite{bal},  and for this
we shall introduce a scalar field $\phi$ on   $\partial \Sigma$.  In terms
of this field,  chirality will mean the following:
\be
 {\cal D}_-\phi= f(x^-) \;,\label{boschi}
\ee
where $x^-={1\over\sqrt{2}}
(x^0-x^1)$, ${\cal D}_\pm={1\over\sqrt{2}}({\cal D}_0\pm{\cal D}_1)$ and
$ {\cal D}_{\mu}$ denotes a covariant derivative.
(Usually the more restrictive condition $f(x^-)=0$ is assumed, but  
Eq.~\ref{boschi} seems enough for us.) 

To proceed  we shall  pose an action principle for the edge field $\phi$.
The edge action  ${\cal S}_{\partial
\Sigma\times R^1}$   should be such that:
i) The total action  ${\cal S}=
{\cal S}_H+{\cal S}_{\partial \Sigma\times R^1}$ is gauge invariant.
ii) It is consistent with chirality, i.e. Eq.~\ref{boschi}.
We will show that these two conditions lead to a chiral electromagnetic
 current  $ J_-^{em} =0$, which at the boundary
 is defined by
$ J^\mu_{em} = -  \frac{\delta {\cal S}}{\delta A_{\mu}}|_{\partial
\Sigma \times R^1}$ .  Requirements
i) and ii) also lead to the anomaly.  For this recall  that   the one loop
effects responsible for the anomaly in the
fermionic theory  appear at tree level
in the bosonized theory.   Thus we can expect  to recover
the anomaly from the classical equation of motion
for $\phi$ \cite{anomaly}.

We begin by addressing the issue i) of gauge invariance.
If we ignore boundary effects, the bulk action
is separately gauge invariant under
transformations of the electromagnetic potentials $A_\mu$,
\be A_\mu \rightarrow A_\mu + \partial_\mu \Lambda \;,
\label{gtoA}\ee as well as under transformations
 of the
fields $g$,
\be
g \rightarrow
g  \;e^{i \lambda \sigma_3}
\;,
\label{GAUGEM}\ee where both $\Lambda$ and $\lambda$ are functions of
space-time coordinates.
On the other hand,
taking into account   the
 boundary $\partial \Sigma$, one
finds instead that Eq.~\ref{gtoA} gives the surface terms
\be \delta
{\cal S}_H=
-\frac{\sigma_H}{2}\int_{
\partial
\Sigma\times R^1}
 d\Lambda\wedge A +
\frac{e\kappa i}{4\pi}\int_{\partial
\Sigma\times R^1}
 d\Lambda\wedge
 {\rm Tr}\sigma_3g^\dagger dg
\;,\label{blkvar2}
\ee
while gauge invariance under transformations Eq.~\ref{GAUGEM}) persists.
We now specify that under gauge transformations
Eq.~\ref{gtoA}), the edge field $\phi$
transforms according to
\be \phi \rightarrow \phi + e\Lambda \;.\label{gtpeL}\ee
Then
we can cancel both of the above boundary terms in
Eq.~\ref{blkvar2} if we assume the following action for the scalar field
$\phi$:
\be
{\cal S}_{\partial
\Sigma\times R^1}
 =\frac{R^2}{8\pi}\int_{\partial
\Sigma\times R^1}
 d^2x\;
({\cal D}_\mu\phi)^2
+ \frac{\sigma_H}{2e}\int_{\partial
\Sigma\times R^1}
 d\phi\wedge A
-\frac{\kappa i}{4\pi}\int_{\partial
\Sigma\times R^1}
 d\phi
\wedge
 {\rm Tr}\sigma_3g^\dagger dg
 \;.
\label{stbdac2}
\ee
In (\ref{stbdac2})
 we have added a kinetic energy term for $\phi$,
where the covariant derivative is  defined by
${\cal D}_\mu\phi=
\partial_\mu\phi  -eA_\mu $.  The coefficient $R$ is real and is known to
correspond to the square root of the filling fraction $\nu_k$. 

Concerning ii), extremizing Eq.~\ref{stbdac2}) with respect to $\phi$ gives
\be\label{aneq2}
R^2 \partial_\mu {\cal  D}^\mu \phi = -\frac{2\pi\sigma_H}{e}F_{01}
-4\pi{\cal T}^r
\;,\ee   $F_{01}$ being the electric
field strength at the boundary and the index $r$   denoting the
direction normal to the surface.  This equation can be
rewritten as
\be\label{aneq20}2
R^2 \partial_+ {\cal  D}_- \phi =(eR^2 -\frac{2\pi\sigma_H}{e})F_{01}
-4\pi{\cal T}^r
\;,\ee using $\partial_+={1\over\sqrt{2}}(\partial_0+\partial_1)$
 and  $diag(1,-1)$ for the Lorenz metric.  But the chirality condition
Eq.~\ref{boschi}) requires that the right hand side of
Eq.~\ref{aneq20}) vanishes.  For this we can set
\be \sigma_H=\frac{e^2 R^2}{2\pi} \;, \label{Rsqnu2}\ee
 which is the  usual relation for the Hall conductivity (after  identifying
 $R^2$
with the filling fraction $\nu_k$).  But we also need
\be
{\cal T}^r = 0 \quad {\rm  at }
\;\partial \Sigma \;.\label{tccp}
\ee
From  Eq.~\ref{Rsqnu2}) and Eq.~\ref{tccp}, variations of $A_\mu$ give the
following result for the edge current
\be\label{bos_curr_el}
    J^{\mu}_{em} = -  \frac{\delta {\cal S}}{\delta A_{\mu}}|_{\partial
\Sigma \times R^1}=\frac{e   R^2}{4\pi}
\left( {\cal D}^{\mu}+\epsilon^{\mu \nu}{\cal D}_{\nu} \right)\phi\;,\ee
  and thus it is chiral, i.e. $ J_-^{em} =0$.  Here
  $\epsilon^{01}=-\epsilon^{10}=1$.  By taking its divergence we also recover
the anomaly:
\be \partial_\mu J^\mu_{em}  =
    \frac{e   R^2}{4\pi} \partial_\mu
\left( {\cal D}^{\mu}+\epsilon^{\mu \nu}{\cal D}_{\nu} \right) \phi
 =-\frac{e^2 R^2}{2\pi} F_{01}
\;, \label{dvj} 
\ee 
where we again used Eq.~\ref{Rsqnu2} and Eq.~\ref{tccp}.

In order to satisfy chirality in the above discussion,
 we needed not only to constrain the values of coefficients, but we also
found it necessary to impose a boundary condition
Eq.~\ref{tccp} on the topological current.  As a result, the topological
flux, and moreover Skyrmions, cannot  penetrate
the edge.  Thus, provided $g$ is everywhere defined in $\Sigma$, the total
Skyrmion number within the bulk
$\int_{\Sigma} d^2x \; T^{0}(g)$
is a conserved quantity, and for example, a nonzero value for the total
topological charge
cannot be adiabatically  generated from the ground state.

Below, we generalize to the  situation where the total
Skyrmion number in the bulk is ${\it not}$
 restricted to being a constant.  For this we need to drop
 the boundary  condition Eq.~\ref{tccp}, and thus allow for
a nonzero topological flux into or out of the sample.
One may interpret this as Skyrmions
being created or destroyed at the edges.
For this purpose, we consider an extension of the above
description, where  the Hopf term
\be
{\cal S}_{WZ} =\frac{\Theta}{24 \pi^2}
\int_{\Sigma\times R^1}{\rm Tr}( g^{\dagger} dg)^3 \label{WZterm}
\ee                is added
 to the bulk action ${\cal S}_H$.
 [Note that
Eq.~\ref{WZterm} is a local version of the Hopf term]. This term does not 
affect the classical equations of motion since it is the integral of a 
closed three form. That term gives a nontrivial spin to the Skyrmion
\be\frac{\Theta N_{Sky}}{2\pi}\;.
\label{spin}
\ee

We thus need the numerical value of $\Theta$ to determine the spin.  For
this purpose we now reexamine the boundary dynamics taking into
account the  Hopf term.  We once again require i) gauge invariance and
ii) chirality.

Concerning i),
  as before, the bulk action is not invariant under
  gauge transformations Eq.~\ref{gtoA}
  of the electromagnetic potentials $A_\mu$.
In addition, unlike before, it is  not invariant under gauge
transformations Eq.~\ref{GAUGEM} of the fields $g$.
From ${\cal S}_{WZ}$ we pick up  the surface term
\be \delta{\cal S}_{WZ}=
\frac{i\Theta}{8\pi^2}\int_{\partial\Sigma\times R^1} d\lambda\wedge
 {\rm Tr}\sigma_3g^\dagger dg\;.\label{blkvar3}\ee
To cancel this variation along with Eq.~\ref{blkvar2}, we once again assume
the existence of an edge field $\phi$ which transforms like Eq.~\ref{gtpeL},
simultaneously with the  gauge transformations Eq.~\ref{gtoA}
  of the electromagnetic potentials $A_\mu$.
We further specify that $\phi$
transforms according to
\be \phi \rightarrow \phi + \lambda
\;,\label{gtsf}
\ee simultaneously
with the gauge transformations Eq.~\ref{GAUGEM} of the fields $g$.
Recall that this transformation is the one compatible with the 
covariant derivative in Eq.~\ref{lag_CSLG}.
Then we can cancel both of the  boundary terms  Eq.~\ref{blkvar2} and
Eq.~\ref{blkvar3}, making our theory anomaly free,
if we assume the following action for the scalar field $\phi$:
  \begin{eqnarray}\label{stbdac5}
{\cal S}_{\partial\Sigma\times R^1}
 &=&\frac{R^2}{8\pi}\int_{\partial\Sigma\times R^1}
 d^2x\;({\cal D}_\mu\phi)^2+ \frac{\sigma_H}{2e}
\int_{\partial\Sigma\times R^1}
 d\phi\wedge A \\
& &- \frac{i}{4}\int_{\partial\Sigma\times R^1}
\biggl(\frac{\Theta}{2\pi^2} d\phi
+\frac{\sigma_H}{e} A \biggr)\wedge {\rm Tr}\sigma_3 g^\dagger dg
\nonumber \;,
\end{eqnarray}
provided we also impose that \be
\kappa  = \frac{\pi\sigma_H}{e^2}+\frac{\Theta}{2\pi}
 \;.\label{formfk}\ee   Since
$\phi$ admits gauge transformations Eq.~\ref{gtsf}, as well as
Eq.~\ref{gtpeL}, we must
redefine the covariant derivative appearing in Eq.~\ref{stbdac5} according
to
\be    {\cal D}_\mu\phi=
\partial_\mu\phi  -\beta_\mu\;,\qquad \beta_\mu  =
eA_\mu -\frac{i}{2}{\rm Tr}\sigma_3 g^\dagger
\partial_\mu g\;.\ee

With regard  to ii),
 the equation of motion for $\phi$ is  \be
R^2 \partial_\mu {\cal  D}^\mu \phi = -\frac{2\pi\sigma_H}{e}F_{01}-
2{\Theta} T^r\;, \ee
which can be rewritten as
\be\label{aneq21}2
R^2 \partial_+ {\cal  D}_- \phi =(eR^2 -\frac{2\pi\sigma_H}{e})F_{01}
+2(\pi R^2-\Theta){\cal T}^r
\;.\ee
  We recover the chirality condition
Eq.~\ref{boschi} upon setting
\be\Theta
=\pi R^2\;,\label{Rsqnu5}
\ee as well as (\ref{Rsqnu2}).
From Eq.~\ref{Rsqnu2}) and Eq.~\ref{Rsqnu5}), variations of $A_\mu$ i
again give the edge current in the form of
Eq.~\ref{bos_curr_el} (although the covariant derivative is now
 defined differently)
 and thus  $ J_-^{em} =0$.   By taking its divergence we get
the anomaly equation:
\be \partial_\mu J^\mu_{em} =-\frac{eR^2}{2\pi}
\epsilon^{\mu\nu} \partial_\mu\beta_\nu
\;. \ee

  Thus now we can satisfy the criterion of chirality
 without imposing any boundary conditions on the topological current.
     Substituting Eq.~\ref{Rsqnu5})  into
 Eq.~\ref{formfk} (and using $R^2=\nu_k$)
  also fixes $\kappa$ to be the filling fraction.  It follows that the
Skyrmion charge is $e\nu_k N_{Sky}$.
 Eqs.~\ref{spin}, \ref{Rsqnu5}
then give the value for the spin to be $ \frac{N_{Sky}
\nu_k}2 $.  Therefore, within the above assumptions, a winding
number one Skyrmion is a fermion when the filling fraction is one.
\bigskip

{\bf Acknowledgments}

It is a pleasure to thank S. Baez, A.P. Balachandran and A. Stern for
so many discussions. I also acknowledge interest and discussions with 
J. Soto and R. Ray.
This research was supported by the U.S. Department of
Energy under contract DE-FG02-85ER40237.


\begin{thebibliography}{99}

\bibitem{WE}
S. Baez, A.P. Balachandran, A. Stern and A. Travesset, {\em cond-mat 9712151}.

\bibitem{Review} R.E. Prange and S.M. Girvin, {\em The Quantum Hall
effect}, Springer-Verlag, New York (1990) and references therein;

\bibitem{zhkl} S. C. Zhang, H. Hansson and S. Kivelson, Phys. Rev. Lett.
{\bf 62}, 82 (1989);
{\bf 62}, 980 (1989);
D. H. Lee and S. C. Zhang,  Phys. Rev. Lett.
{\bf 66}, 1220 (1991).

\bibitem{SHOU} 
S. C. Zhang,  Int. J. of Mod. Phys. {\bf B 6}, 25 (1992) and references 
therein.

\bibitem{NORW} J.M. Leinaas and S. Viefers, {\em cond-mat 9712009}.

\bibitem{soto}
R. Ray, This proceedings.

\bibitem{Sondhi} S.L. Sondhi, A. Karlhede, S.A. Kivelson and E.H. Rezayi,
Phys. Rev. {\bf B 47}, 16419 (1993).

\bibitem{Fro} X.Wen, Adv. In Phys.\ 44 (1995) 405-473 and references therein.

\bibitem{bal} A.P. Balachandran, L. Chandar and B. Sathiapalan,
Nucl. Phys. {\bf B 443}, 465 (1996); Int. J. of Mod. Phys. {\bf A 11},
3587 (1996).

\bibitem{anomaly}
C.G. Callan Jr. and J.A. Harvey, Nucl. Phys. B 250 (1985) 427;
S.G. Naculich, Nucl. Phys. B 296 (1988) 837.

\end{thebibliography}
\end{document}